\documentclass[reprint,amsmath,amssymb,aps,pra,showpacs]{revtex4-1}
\usepackage{bm}
\usepackage{graphicx,graphics}
\usepackage{dcolumn}
\usepackage{hyperref}

\begin{document}

\title{Family of Finite Geometry Low-Density Parity-Check Codes for Quantum Key Expansion}
\author{Kung-Chuan Hsu}
  \email{kungchuh@usc.edu}
\author{Todd A. Brun}
  \email{tbrun@usc.edu}
\affiliation{Ming Hsieh Department of Electrical Engineering, University of Southern California, Los Angeles, California 90089, USA}
\date{\today}

\begin{abstract}
We consider a quantum key expansion (QKE) protocol based on entanglement-assisted quantum error-correcting codes (EAQECCs). In these protocols, a seed of a previously shared secret key is used in the post-processing stage of a standard quantum key distribution protocol like the Bennett-Brassard 1984 protocol, in order to produce a larger secret key. This protocol was proposed by Luo and Devetak, but codes leading to good performance have not been investigated. We look into a family of EAQECCs generated by classical finite geometry (FG) low-density parity-check (LDPC) codes, for which very efficient iterative decoders exist. A critical observation is that almost all errors in the resulting secret key result from uncorrectable block errors that can be detected by an additional syndrome check and an additional sampling step. Bad blocks can then be discarded. We make some changes to the original protocol to avoid the consumption of the preshared key when the protocol fails.  This allows us to greatly reduce the bit error rate of the key at the cost of a minor reduction in the key production rate, but without increasing the consumption rate of the preshared key. We present numerical simulations for the family of FG LDPC codes, and show that this improved QKE protocol has a good net key production rate even at relatively high error rates, for appropriate choices of these codes.
\end{abstract}


\maketitle

\section{Introduction}
\label{I}
A quantum key expansion (QKE) protocol allows two parties, Alice and Bob, to expand a shared secret key by using one-way quantum communication and public classical communication. Luo and Devetak \cite{Luo07} demonstrated a QKE protocol, which is derived from the standard Bennett-Brassard 1984 (BB84) quantum key distribution (QKD) protocol with post-processing steps involving the use of entanglement-assisted Calderbank-Shor-Steane (CSS) codes. The protocol is provably secure from an eavesdropper, Eve, based on a result by Shor and Preskill \cite{Shor00}.

The QKE protocol has a potential advantage over QKD, in that the original pair of classical codes considered need not have the dual-containing property. The cost is that the parties involved have to pre-share a secret key. The classical codes correspond to entanglement-assisted quantum error-correcting codes (EAQECCs). The EAQECC construction is described by the formalism given by Brun, Devetak, and Hsieh \cite{Brun06}.

In the CSS construction of Luo and Devetak's QKE protocol, a pair of classical linear codes with good error-correcting performance is needed. Low-density parity-check (LDPC) codes are classical linear codes that have sparse parity-check matrices, and many families of LDPC codes have been studied and claimed to give good performance (see, e.g., \cite{Kou01,MacKay04,Camara05,Hagiwara07,Poulin08,Hsieh09,Hsieh11}). There have been several recent studies on the performance of LDPC codes used for QKD \cite{Ohata07,Elkouss09}. In this paper, LDPC codes constructed from finite geometry (FG) are considered \cite{Kou01,Hsieh11}, and methods to incorporate them into the QKE protocol are proposed and explained. For simplicity, the quantum channel is modeled by a depolarizing channel. Given a tolerable bit error threshold $\epsilon$ for the generated keys, the goal is to search for codes that maximize the net key rate for given channel error parameters.

The paper is organized as follows. In Sec.~\ref{II}, we first introduce the QKE protocol of Luo and Devetak. We then propose modifications to the post-processing steps to improve performance. In Sec.~\ref{III}, we discuss families of LDPC codes generated by finite geometry. In Sec.~\ref{IV}, we discuss simulation results using the improved QKE protocol from Sec.~\ref{II} and the codes from Sec.~\ref{III}, and we analyze their performance. In Sec.~\ref{V}, we give conclusions and suggest possible work in the near future.

The one-dimensional vectors appearing in this paper should always be considered as column vectors. The vectors are denoted with underlined italic characters, and the matrices are denoted with boldface italic characters. The operations $+$ and $\oplus$ are defined respectively as component-wise addition and addition modulo $2$.

\section{Quantum Key Expansion}
\label{II}
The QKE protocol discussed in this paper is derived from the BB84 quantum key distribution protocol, using CSS codes for error correction and privacy amplification. The CSS code used for a BB84 QKD protocol is derived from a pair of ``dual-containing" classical linear codes. Most pairs of classical codes do not satisfy this requirement, but such pairs can be found. The dual-containing property requires that $\bm{H}_1 \bm{H}_2^T=\bm{0}$ be satisfied, where $\bm{H}_1$ and $\bm{H}_2$ are the parity-check matrices of the two codes. The QKE protocol, however, does not require the pair of classical codes to have the dual-containing restriction. The idea is to interpret the code as an entanglement-assisted code rather than a standard quantum code, and the cost is that the two parties involved must have a preshared secret key that is expanded by the protocol.

In Sec.~\ref{IIA}, the structure of entanglement-assisted codes will be introduced, as well as the notation that will be used throughout the paper. Section~\ref{IV} reviews the steps of the QKE protocol proposed by Luo and Devetak \cite{Luo07}. In Sec.~\ref{IIC} and ~\ref{IID}, we analyze the post-processing steps of the QKE protocol and propose improvements. In Sec.~\ref{IIE}, we summarize the improvements of Sec.~\ref{IID} and give a QKE protocol with enhanced performance compared to the original QKE protocol.

\subsection{Code construction}
\label{IIA}
This section summarizes the entanglement-assisted CSS code construction and the matrix structures involved. The notation mentioned here will be used throughout the later sections.

For $i=1,2$, let $C_i$ be a classical $[n,k_i,d]$ code with parity-check matrix $\bm{H}_i$ of size $(n-k_i)\times n$. Based on the given pair of classical codes, an $[[n,k_1+k_2-n+c,d;c]]$ entanglement-assisted quantum CSS code can be constructed, where $c=\mbox{rank}(\bm{H}_1 \bm{H}_2^T)$ is the number of ebits (or entangled pairs of qubits) needed. This code can protect $m=k_1+k_2-n+c$ qubits from error. After this process, we end up with two dual-containing classical codes $C_1^{\prime}$ and $C_2^{\prime}$ with ``augmented" parity check matrices $\bm{H}_1^\prime$ and $\bm{H}_2^\prime$. The derivation of $\bm{H}_i^\prime$ from $\bm{H}_i$ is as follows:

For a given pair of $\bm{H}_1$ and $\bm{H}_2$, there always exist nonsingular matrices $\bm{T}_1$ and $\bm{T}_2$ such that

\begin{equation}
\bm{T}_1 \bm{H}_1 \bm{H}_2^T \bm{T}_2^T=
\left( \begin{array}{cc}
\bm{0}_{(n-k_1-c)\times(n-k_2-c)} & \bm{0}_{(n-k_1-c)\times c} \\
\bm{0}_{c\times (n-k_2-c)} & \bm{I}_c \\
\end{array} \right).
\end{equation}

$\bm{H}_i^\prime$ can thus be constructed as follows to assure that the new codes satisfy the dual-containing property, $\bm{H}_1^\prime \bm{H}_2^{\prime T}=\bm{0}$.

\begin{equation}
\bm{H}_i^\prime=(\bm{T}_i \bm{H}_i~\bm{J}_i)
\text{, where }
\bm{J}_i=
\left( \begin{array}{c}
\bm{0}_{(n-k_i-c)\times c} \\
\bm{I}_c \\
\end{array} \right).
\end{equation}

Suppose $\bm{H}_1^\prime$ and $\bm{H}_2^\prime$ are constructed. There exist binary matrices $\bm{E}_1$, $\bm{F}_1$, $\bm{E}_2$, and $\bm{F}_2$ such that the following four requirements are satisfied:

1. The rows of $\bm{H}_1^\prime$ and $\bm{E}_1$ form a basis for $C_2^\prime$.

2. The rows of $\bm{H}_2^\prime$ and $\bm{E}_2$ form a basis for $C_1^\prime$.

3. $\bm{N}_1=
\left( \begin{array}{c}
\bm{H}_1^\prime \\
\bm{E}_1 \\
\bm{F}_1 \\
\end{array} \right)$
and $\bm{N}_2=
\left( \begin{array}{c}
\bm{F}_2 \\
\bm{E}_2 \\
\bm{H}_2^\prime \\
\end{array} \right)$
are full rank matrices.

4. $\bm{N}_1 \bm{N}_2^T=\bm{I}$.

The new parity-check matrices $\bm{H}_i^\prime$ have more columns than the original $\bm{H}_i$. These columns correspond to additional qubits on the receiver's side. Before decoding, the sender (Alice) and the receiver (Bob) share $c$ entangled pairs. Since Bob's half of these pairs do not pass through the channel, they are noise-free.

The syndrome of an error is defined as the error vector multiplied by the parity-check matrix of the code. For the code $C_1^\prime$ in our case, the syndrome corresponding to the error vector $\underline{e}$ is $\underline{s}=\bm{H}_1^\prime \underline{e}$. The set of codewords of the code is the set of all vectors with zero syndromes.

The decoder for the LDPC codes considered in this paper is a sum-product algorithm (SPA) decoder \cite{MacKay99} that identifies a probable error corresponding to each syndrome. Based on the decoder, the error set correctable by the code can be defined. For the code $C_1^\prime$ with parity-check matrix $\bm{H}_1^\prime$, one may define such a set as $\mathcal{E}_1^\prime=\{\bm{F}_2^T \underline{s}+\bm{E}_2^T\underline{\beta}(\underline{s})+\bm{H}_2^{\prime T}\underline{\beta}^\prime(\underline{s}):\underline{s}\in\mathbb{Z}_2^{n-k_1}\}$, where $\underline{\beta}(\underline{\cdot}):\mathbb{Z}_2^{n-k_1}\rightarrow \mathbb{Z}_2^m$ and $\underline{\beta}^\prime(\underline{\cdot}):\mathbb{Z}_2^{n-k_1}\rightarrow \mathbb{Z}_2^{n-k_2}$ are mappings fixed by the decoder. For every syndrome $\underline{s}\in\mathbb{Z}_2^{n-k_1}$, the decoder gives $\bm{F}_2^T\underline{s}+\bm{E}_2^T\underline{\beta}(\underline{s})+\bm{H}_2^{\prime T}\underline{\beta}^\prime(\underline{s})$ as the probable error. The receiver then corrects this error on the received codeword to retrieve the original message.

\subsection{Luo and Devetak's quantum key expansion protocol}
\label{IIB}
Let Alice and Bob be the sender and receiver utilizing the QKE protocol proposed in \cite{Luo07}. The steps of the protocol are as follows:

1) Alice generates a binary string $\underline{a}$ consisted of $(2+3\delta)n$ random bits.

2) Alice generates another binary string $\underline{\alpha}$ consisted of $(2+3\delta)n$ random bits, and she prepares each bit in $\underline{a}$ in the $Z$ or $X$ basis according to the corresponding bit in $\underline{\alpha}$. For example, Alice may prepare the bit in $\underline{a}$ in the $Z$ basis if the corresponding bit in $\underline{\alpha}$ is $0$, and in the $X$ basis otherwise.

3) Alice sends the prepared qubits to Bob.

4) Bob receives the qubits, and he generates a binary string $\underline{\gamma}$ consisting of $(2+3\delta)n$ random bits. Bob then uses $\underline{\gamma}$ to determine in which bases to measure the received qubits. To be consistent with the example in 2), Bob measures the received qubit in the $Z$ basis if the corresponding bit in $\underline{\gamma}$ is $0$ and measures in the $X$ basis otherwise. Let the resulting bit string be $\underline{b}$.

5) Alice announces $\underline{\alpha}$, and Bob discards the bits in $\underline{b}$ where the corresponding bits in $\underline{\gamma}$ and $\underline{\alpha}$ do not match, that is, the bit locations where they prepare and measure in different bases. Bob announces which bits he discards. With high probability, there are at least $(1+\delta)n$ bits left; if not, they abort and restart the protocol.

6) Alice randomly chooses $n$ bits and announces the bit locations for Bob to extract the corresponding bits. Let Alice's resulting string be $\underline{\hat{a}}$, and Bob's be $\underline{\hat{b}}$. There are at least $n\delta$ pairs of bits left, and those pairs are used for channel estimation. Alice and Bob announce those bits to each other and count the fraction that do not match. If there are too many errors, they abort and restart the protocol.

7) Alice attaches the length-$c$ preshared bit string $\underline{\kappa}$ to $\underline{\hat{a}}$. She first computes $\underline{s}_A=\bm{H}_1^\prime
\left( \begin{array}{c}
\underline{\hat{a}} \\
\underline{\kappa} \\
\end{array} \right)$
and announces it to Bob. She then computes her part of the generated key, $\underline{k_A}=\bm{E_1}
\left( \begin{array}{c}
\underline{\hat{a}} \\
\underline{\kappa} \\
\end{array} \right)$.

8) Bob computes $\underline{s}_B=\bm{H}_1^\prime
\left( \begin{array}{c}
\underline{\hat{b}} \\
\underline{\kappa} \\
\end{array} \right)$, and his part of the generated key is $\underline{k}_B=\bm{E}_1
\left( \begin{array}{c}
\underline{\hat{b}} \\
\underline{\kappa} \\
\end{array} \right)
\oplus\underline{\beta}(\underline{s}_A\oplus \underline{s}_B)$.

\subsection{Analysis of QKE post-processing}
\label{IIC}
Consider the procedure of Luo and Devetak's QKE protocol formalized in the previous section. The error correction is performed at the last step 8) where Bob computes $\underline{\beta}(\underline{s}_A\oplus \underline{s}_B)$. In this case, $\underline{s}_A\oplus \underline{s}_B$ is the syndrome that initializes the decoding. To understand how the function $\underline{\beta}(\underline{\cdot})$ is computed, we need to examine its definition and the matrix structure of the code.

Suppose we start with two LDPC codes with parity-check matrices $\bm{H}_1$ and $\bm{H}_2$ of sizes $(n-k_1)\times n$ and $(n-k_2)\times n$, and $c=\mbox{rank}(\bm{H}_1 \bm{H}_2^T)$. The formalism in Sec.~\ref{IIA} gives two $(n+c)\times(n+c)$ full rank matrices $\bm{N}_1$ and $\bm{N}_2$, each formed by three block-matrices $\bm{H}_i^\prime$, $\bm{E}_i$, and $\bm{F}_i$ of sizes $(n-k_i)\times(n+c), (k_1+k_2-n+c)\times(n+c)$, and $(n-k_{(1+i \mbox{ \scriptsize mod}2)})\times(n+c)$, respectively. $\bm{H}_1^\prime$ and $\bm{H}_2^\prime$ are defined as the parity check matrices of the newly formed entanglement-assisted CSS code. Note that the two new parity-check matrices need not be low-density and thus the performance will be poor if one uses them to run the SPA decoder. However, as seen in Sec.~\ref{IIA}, since the matrix operations transforming $\bm{H}_i$ to $\bm{H}_i^\prime$ are reversible, the error syndrome with respect to the original parity-check matrix $\bm{H}_i$ can be retrieved by doing inverse matrix operations on the corresponding syndrome with respect to $\bm{H}_i^\prime$. That is, given a syndrome corresponding to $\bm{H}_i^\prime$, we can find the corresponding syndrome for $\bm{H}_i$. As a result, the errors can be decoded by the SPA decoder with LDPC matrix $\bm{H}_i$. The details follow.

The function $\underline{\beta}(\underline{\cdot})$, which includes the process of error correction, comes into the picture when the error set $\mathcal{E}_1$ correctable by the code $\bm{H}_1^\prime$ is defined. Recall from Sec.~\ref{IIA}, that $\mathcal{E}_1=\{\bm{F}_2^T\underline{s}+\bm{E}_2^T\underline{\beta}(\underline{s})+\bm{H}_2^{\prime T}\underline{\beta}^\prime(\underline{s}):\underline{s}\in\mathbb{Z}_2^{n-k_1}\}$. Since the matrix $\bm{N}_2$ formed by $\bm{H}_2^\prime$, $\bm{E}_2$, and $\bm{F}_2$ is a full rank matrix in $\mathbb{Z}_2$, the error string corresponding to a particular syndrome $\underline{s}$ can be retrieved by the following steps:\\

i) Compute $\underline{s}^\prime=\bm{T}_1^{-1}\underline{s}$.\\

ii) Run the SPA decoder using the original LDPC matrix $\bm{H}_1$ with the syndrome $\underline{s}^\prime$. The decoded string is the estimated error, and we denote it by $\underline{\hat{e}}$.\\

iii) Attach $c$ $0$'s to $\underline{\hat{e}}$ and compute $
\underline{\beta}(\underline{s})=\bm{E}_1
\left( \begin{array}{c}
\underline{\hat{e}} \\
\underline{0}_{c\times 1} \\
\end{array} \right)$.

In the above steps i) and ii), the error message can be decoded using $\bm{H}_1$ instead of $\bm{H}_1^\prime$ since the last $c$ bits of the message are preshared by Alice and Bob, and thus the error message from those bits should always be a string of $0$'s. The syndrome is then totally determined by the first $n$ bits of the error message. This allows us to use the original low-density parity-check matrices for decoding and thus the error-correcting performance is maintained.

The last step may not be trivial, and we explain it in the following. Using our notation, if $
\left( \begin{array}{c}
\underline{\hat{e}} \\
\underline{0}_{c\times 1} \\
\end{array} \right)$ is correctable by $\bm{H}_1^\prime$ with syndrome $\underline{s}$, it is in the set $\mathcal{E}_1$ and can be written in the form

\begin{equation}
\left( \begin{array}{c}
\underline{\hat{e}} \\
\underline{0}_{c\times 1} \\
\end{array} \right)=\bm{N}_2^T
\left( \begin{array}{c}
\underline{s} \\
\underline{\beta}(\underline{s}) \\
\underline{\beta}^\prime(\underline{s}) \\
\end{array} \right).
\end{equation}

Since $\bm{N}_1 \bm{N}_2^T=\bm{I}$, it is obvious that $\bm{N}_2^T=\bm{N}_1^{-1}$. $\bm{N}_1$ can then be multiplied to both sides of the above equation. As a result,

\begin{equation}
\left( \begin{array}{c}
\underline{s} \\
\underline{\beta}(\underline{s}) \\
\underline{\beta}^\prime(\underline{s}) \\
\end{array} \right)=\bm{N}_1
\left( \begin{array}{c}
\underline{\hat{e}} \\
\underline{0}_{c\times 1} \\
\end{array} \right)=
\left( \begin{array}{c}
\bm{H}_1^\prime \\
\bm{E}_1 \\
\bm{F}_1 \\
\end{array} \right)
\left( \begin{array}{c}
\underline{\hat{e}} \\
\underline{0}_{c\times 1} \\
\end{array} \right).
\end{equation}

It should now be clear that step iii) is valid.

\subsection{Improving QKE post-processing}
\label{IID}
A very important observation based on our simulations is that in the cases where the channel error rates are not small, the bit error rates of the resulting keys are significant whenever the estimated errors $
\left( \begin{array}{c}
\underline{\hat{e}} \\
\underline{0}_{c\times 1} \\
\end{array} \right)$ are erroneous. Specifically, the bit error rates of the keys are about half the block error rates for sufficiently large channel error probabilities. Since $\underline{\beta}(\underline{\cdot})$ is equivalent to multiplying by a matrix, $\bm{E}_1$, this observation implies that $\bm{E}_1$ is generally not sparse. Given a block error, it is likely that each row of $\bm{E}_1$ and the block error have overlapping non-zero elements, which on average contributes to a significant number of errors in the key. In other words, when a block error occurs the resulting key is almost totally randomized.

From the observation above, we can apply two useful improvements to the protocol.

$Improvement$ $1.$ This is to check the syndrome following the decoder's output. This allows the detection of not-yet-converged messages from the SPA decoder. These messages must have block errors. Aborting the protocol after detecting those erroneous messages greatly improves the error performance of the generated key, at the cost of modestly reducing the key rate, since the information sent through the channel in the prior stages is wasted.

$Improvement$ $2.$ This is to check the generated keys directly. Let the block error rate and bit error rate of the generated keys be denoted by $R_{blk}$ and $R_{bit}$. Since block errors of the keys result in a large fraction of the bits being erroneous in each block, checking several randomly chosen bits allows a large probability of detecting those block errors. Let us assume the relationship $R_{bit}=q R_{blk}$, such that, on average, a block error yields a bit error rate of $q$. Suppose each time the protocol is processed, a number of bits $\mu$ are chosen randomly from the key, and are used for a check between the sender and the receiver. The bit error rate of the generated key, $\hat{R}_{bit}$, can then be calculated as

\begin{equation}
\hat{R}_{bit}=R_{bit}\frac{\left(
1-q\right)
^{\mu}}{1-R_{blk}+\left(
1-q\right)
^{\mu}R_{blk}}\equiv
R_{bit} f.
\end{equation}

The bit error rate is scaled by the factor $f$.
For fixed $R_{blk}$, $f$ decreases dramatically as $\mu$ increases. This means that not many bits need be checked to greatly improve the error performance of the key. To determine $\mu$, we find the smallest $\mu$ satisfying $\hat{R}_{bit}<\epsilon$, where $\epsilon$ is the desired threshold for the bit error rate of the final key. That is,

\begin{equation}
\mu=
\left\{
\begin{array}{cl}
\lceil log_{(1-q)}(\frac{\epsilon(1-R_{blk})}{(q-\epsilon)R_{blk}})\rceil & \mbox{ if } q>\epsilon\mbox{,} \\
0 & \mbox{ otherwise.} \\
\end{array}
\right.
\end{equation}

Since those randomly chosen $\mu$ bits from the key are revealed, the tradeoff in using this method would be to reduce the key rate by an amount $\frac{\mu}{n}$.

A problem arises here, in that the preshared key bits are consumed even if the protocol fails, which could even result in the net key rate being negative. However, there is a way to get around this problem.

In the original QKE protocol, Alice announces to Bob the message $\underline{s}_A=\bm{H}_1^\prime
\left( \begin{array}{c}
\underline{\hat{a}} \\
\underline{\kappa} \\
\end{array} \right)$, and Bob corrects the errors using the syndrome $\underline{s}=\underline{s}_A\oplus \bm{H}_1^\prime
\left( \begin{array}{c}
\underline{\hat{b}} \\
\underline{\kappa} \\
\end{array} \right)=\bm{H}_1^\prime
\left( \begin{array}{c}
\underline{\hat{a}}\oplus\underline{\hat{b}} \\
\underline{0} \\
\end{array} \right)$. This syndrome can also be computed by Bob if Alice sends the message $\underline{\hat{s}}_A=\bm{H}_1^\prime
\left( \begin{array}{c}
\underline{\hat{a}} \\
\underline{0} \\
\end{array} \right)$ instead. In this case, Bob just computes $\underline{s}=\underline{\hat{s}}_A\oplus \bm{H}_1^\prime
\left( \begin{array}{c}
\underline{\hat{b}} \\
\underline{0} \\
\end{array} \right)=\bm{H}_1^\prime
\left( \begin{array}{c}
\underline{\hat{a}}\oplus\underline{\hat{b}} \\
\underline{0} \\
\end{array} \right)$.

Thus, instead of comparing the keys $\underline{k}_A=\bm{E}_1
\left( \begin{array}{c}
\underline{\hat{a}} \\
\underline{\kappa} \\
\end{array} \right)$ and $\underline{k}_B=\bm{E}_1
\left( \begin{array}{c}
\underline{\hat{b}} \\
\underline{\kappa} \\
\end{array} \right)
\oplus\underline{\beta}(\underline{s}_A\oplus \underline{s}_B)$ and consuming the preshared key $\underline{\kappa}$, it is sufficient for the two parties to compare $\underline{\hat{k}}_A=\bm{E}_1
\left( \begin{array}{c}
\underline{\hat{a}} \\
\underline{0} \\
\end{array} \right)$ and $\underline{\hat{k}}_B=\bm{E}_1
\left( \begin{array}{c}
\underline{\hat{b}} \\
\underline{0} \\
\end{array} \right)
\oplus\underline{\beta}(\underline{\hat{s}}_A\oplus\underline{\hat{s}}_B)$.
In this way, we can postpone the consumption of the preshared keys until after the check is performed. Note that, Alice and Bob must discard the bits from the final key corresponding to the ones they compare, since information about those bits is publicly revealed.

\subsection{Summary of the improved QKE protocol}
\label{IIE}
In this section, we will combine the two improvements from the previous section and assess the improved performance of the QKE protocol. We consider the case where Improvement 1 is performed first, and then Improvement 2 is performed if the check in Improvement 1 is successful.

Let $p_1$ be the failure rate of the check in Improvement 1. Conditioned on passing the check in Improvement 1, let $p_2$ be the rate of bit errors in the generated keys followed by the remaining block errors. Also, let $R_{blk}$ be the block error rate of the LDPC code and $\epsilon$ be the error threshold that is desired for QKE. The values, $R_{blk}$, $p_1$ and $p_2$, can be determined by simulation. After Improvement 2 is performed, the bit error rate of the generated key, $\hat{R}_{bit}$, can then be calculated:

\begin{equation}
\hat{R}_{bit}=p_2\frac{\left(
1-p_2\right)
^{\mu}\left(
R_{blk}-p_1\right)}{
1-R_{blk}+\left(
1-p_2\right)
^{\mu}\left(
R_{blk}-p_1\right)}.
\end{equation}

To determine $\mu$, we find the smallest $\mu$ satisfying $\hat{R}_{bit}<\epsilon$. That is,

\begin{equation}
\mu=
\left\{
\begin{array}{cl}
\lceil log_{(1-p_2)}(\frac{\epsilon(1-R_{blk})}{(p_2-\epsilon)(R_{blk}-p_1)})\rceil & \mbox{ if } p_2>\epsilon\mbox{,} \\
0 & \mbox{ otherwise.} \\
\end{array}
\right.
\end{equation}.

We now outline the improved QKE protocol. Referring to the original QKE protocol in Sec.~\ref{IIB}, the procedure up to step 6) will be the same. The steps beyond 7) are modified as follows:

7) Alice computes $\underline{\hat{s}}_A=\bm{H}_1^\prime
\left( \begin{array}{c}
\underline{\hat{a}} \\
\underline{0} \\
\end{array} \right)$ and announces it to Bob.

8) Bob first computes $\underline{\hat{s}}_B=\bm{H}_1^\prime
\left( \begin{array}{c}
\underline{\hat{b}} \\
\underline{0} \\
\end{array} \right)$, and then he runs the SPA decoder using the original LDPC matrix $\bm{H}_1$ with the syndrome $\underline{s}^\prime=\bm{T}_1^{-1} (\underline{\hat{s}}_A\oplus\underline{\hat{s}}_B)$. Let the decoded error string be $\underline{\hat{e}}$.

9) Bob checks if $\bm{H}_1 \underline{\hat{e}}\oplus \underline{s}^\prime$ is the all-zero string. If not, the protocol is aborted and they start over. This is a result of Improvement 1.

10) Alice randomly chooses $\mu$ bits from $\underline{\hat{k}}_A=\bm{E}_1
\left( \begin{array}{c}
\underline{\hat{a}} \\
\underline{0} \\
\end{array} \right)$ and announces them to Bob. Bob checks if the corresponding bits from $\underline{\hat{k}}_B=\bm{E}_1
\left( \begin{array}{c}
\underline{\hat{b}}\oplus\underline{\hat{e}} \\
\underline{0} \\
\end{array} \right)$ match the ones sent by Alice. If the strings do not completely match, the protocol is aborted and they start over. This is a result of Improvement 2.

11) Alice computes her part of the generated key as $\underline{k}_A=\underline{\hat{k}}_A\oplus \bm{E}_1
\left( \begin{array}{c}
\underline{0} \\
\underline{\kappa} \\
\end{array} \right)$, excluding the $\mu$ bits corresponding to the ones they have compared in the previous step. Bob also computes his part of the generated key as $\underline{k}_B=\underline{\hat{k}}_B\oplus \bm{E}_1
\left( \begin{array}{c}
\underline{0} \\
\underline{\kappa} \\
\end{array} \right)$, excluding the $\mu$ bits similarly.

The preshared key is only used in the last step. Therefore, the preshared key will not be consumed if the protocol is aborted in step 10) or 11). The net key rate of this improved QKE protocol is

\begin{equation}
R_{net}=\left(
1-R_{blk}+\left(
1-p_2\right)
^\mu\left(
R_{blk}-p_1\right)
\right)\frac{m-c-\mu}{n}.
\end{equation}

We will see how well this does in simulations below.

\section{Finite Geometry LDPC Codes}
\label{III}
Finite geometry LDPC codes were formalized by Kou, Lin and Fossorier \cite{Kou01}. There are four families of FG LDPC codes: type-1 Euclidean geometry (EG1) LDPC codes, type-2 Euclidean geometry (EG2) LDPC codes, type-1 projective geometry (PG1) LDPC codes, and type-2 projective geometry (PG2) LDPC codes.  These classical FG LDPC codes were used by Hsieh, Yen and Hsu to construct EAQECCs with good performance that use relatively little entanglement \cite{Hsieh11}.  In this section, we briefly restate the results from \cite{Kou01} and \cite{Hsieh11} and introduce the construction of FG LDPC codes.

\subsection{Euclidean geometry (EG) LDPC codes}
\label{IIIA}
Let EG$(p,2^s)$ be a $p$-dimensional Euclidean geometry over the Galois field GF$(2^s)$, where $p,s\in\mathbb{N}$. This geometry consists of $2^{ps}$ points, where each is a $p$-tuple over GF$(2^s)$. The all-zero $p$-tuple is defined as the origin. Those points form a $p$-dimensional vector space over GF$(2^s)$. A line in EG$(p,2^s)$ is a coset of a one-dimensional subspace of EG$(p,2^s)$, and each line consists of $2^s$ points. There are $2^{(p-1)s}(2^{ps}-1)/(2^s-1)$ lines. Each line has $2^{(p-1)s}-1$ lines parallel to it. Each point is intersected by $(2^{ps}-1)/(2^s-1)$ lines.

Let GF$(2^{ps})$ be the extension field of GF$(2^s)$. Each element in GF$(2^{ps})$ can be represented as a $p$-tuple over GF$(2^s)$, and hence a point in EG$(p,2^s)$. Therefore, GF$(2^{ps})$ may be regarded as the Euclidean geometry EG$(p,2^s)$. Let $\alpha$ be a primitive element of GF$(2^{ps})$. Then $0,\alpha^0,\alpha^1,\alpha^1,...,\alpha^{2^{ps}-2}$ represent the $2^{ps}$ points of EG$(p,2^s)$.

Let $\bm{H}_{EG1}(p,s)$ be a matrix over GF$(2)$. The rows of $\bm{H}_{EG1}(p,s)$ are the incidence vectors of all the lines in EG$(p,2^s)$ not passing through the origin. The columns of $\bm{H}_{EG1}(p,s)$ are the $2^{ps}-1$ non-origin points of EG$(p,2^s)$, and the $i$th column corresponds to the point $\alpha^{i-1}$. Then $\bm{H}_{EG1}(p,s)$ consists of $n=2^{ps}-1$ columns and $J=(2^{(p-1)s}-1)(2^{ps}-1)/(2^s-1)$ rows, and it has the following structure:

1. Each row has weight $\rho_r=2^s$.

2. Each column has weight $\rho_c=(2^{ps}-1)/(2^s-1)-1$.

3. Any two columns have at most one 1-component in common.

4. Any two rows have at most one 1-component in common.

The density of $\bm{H}_{EG1}(p,s)$ is $2^s/(2^{ps}-1)$, which is small for $p$ or $s$ large. Then $\bm{H}_{EG1}(p,s)$ is a low-density matrix.

The LDPC code with parity-check matrix $\bm{H}_{EG1}(p,s)$ is called a type-1 Euclidean geometry LDPC code, and we denote it by $EG1(p,s)$.

Let $\bm{H}_{EG2}(p,s)=\bm{H}_{EG1}(p,s)^T$. Then $\bm{H}_{EG2}(p,s)$ is a matrix with $2^{ps}-1$ rows and $(2^{(p-1)s}-1)(2^{ps}-1)/(2^s-1)$ columns. The rows of $\bm{H}_{EG2}(p,s)$ are the non-origin points of EG$(p,2^s)$, and the columns are the lines in EG$(p,2^s)$ not passing through the origin, and it has the following structure:

1. Each row has weight $\rho_r=(2^{ps}-1)/(2^s-1)-1$.

2. Each column has weight $\rho_c=2^s$.

3. Any two columns have at most one 1-component in common.

4. Any two rows have at most one 1-component in common.

The LDPC code with parity-check matrix $\bm{H}_{EG2}(p,s)$ is called a type-2 Euclidean geometry LDPC code, and we denote it by $EG2(p,s)$.

\subsection{Projective geometry (PG) LDPC codes}
\label{IIIB}
Let GF$(2^{(p+1)s})$ be the extension field of GF$(2^s)$. Let $\alpha$ be a primitive element of GF$(2^{(p+1)s})$. Let $n=(2^{(p+1)s}-1)/(2^s-1)$ and $\eta=\alpha^n$. Then $\eta$ has order $2^s-1$, and the $2^s$ elements $0,\eta^0,\eta^1,\eta^2,...,\eta^{2^s-2}$ form all the elements of GF$(2^s)$. Consider the set $\{\alpha^0,\alpha^1,\alpha^2,...,\alpha^{n-1}\}$, and partition the non-zero elements of GF$(2^{(m+1)s})$ into $n$ disjoint subsets $\{\alpha^i,\eta\alpha^i,\eta^2\alpha^i,...,\eta^{2^s-2}\alpha^i\}$, for $i\in\{0,1,...,n-1\}$. Each such set is represented by its first element $(\alpha^i)$, for $i\in\{0,1,...,n-1\}$.

If each element in GF$(2^{(p+1)s})$ is represented as a $(p+1)$-tuple over GF$(2^s)$, then $(\alpha^i)$ consists of $2^s-1$ $(p+1)$-tuples over GF$(2^s)$. The $(p+1)$-tuple over GF$(2^s)$ that represents $(\alpha^i)$ can be regarded as a point in a finite geometry over GF$(2^s)$. Then the points $(\alpha^0),(\alpha^1),(\alpha^2),...,(\alpha^{n-1})$ form a $p$-dimensional projective geometry over GF$(2^s)$, denoted PG$(p,2^s)$. (Note that a projective geometry does not have an origin.)

Let $\bm{H}_{PG1}(p,s)$ be a matrix over GF$(2)$. The rows of $\bm{H}_{PG1}(p,s)$ are the incidence vectors of all the lines in PG$(p,2^s)$. The columns of $\bm{H}_{PG1}(p,s)$ are the $n$ points of PG$(p,2^s)$, and the $i$th column corresponds to the point $(\alpha^{i-1})$. Then $\bm{H}_{PG1}(p,s)$ consists of $n=(2^{(p+1)s}-1)/(2^s-1)$ columns and $J=(2^{ps}+...+2^s+1)(2^{(p-1)s}+...+2^s+1)/(2^s+1)$ rows, and it has the following structure:

1. Each row has weight $\rho_r=2^s+1$.

2. Each column has weight $\rho_c=(2^{ps}-1)/(2^s-1)$.

3. Any two columns have at most one 1-component in common.

4. Any two rows have at most one 1-component in common.

The density of $\bm{H}_{PG1}(p,s)$ is $(2^{2s}-1)/(2^{(p+1)s}-1)$, which is small for $p$ or $s$ large. Then $\bm{H}_{PG1}(p,s)$ is a low-density matrix.

The LDPC code with parity-check matrix $\bm{H}_{PG1}(p,s)$ is called a type-1 projective geometry LDPC code, and we denote it by $PG1(p,s)$.

Let $\bm{H}_{PG2}(p,s)=\bm{H}_{PG1}(p,s)^T$. Then $\bm{H}_{PG2}(p,s)$ is a matrix with $(2^{(p+1)s}-1)/(2^s-1)$ rows and $(2^{ps}+...+2^s+1)(2^{(p-1)s}+...+2^s+1)/(2^s+1)$ columns. The rows of $\bm{H}_{PG2}(p,s)$ are the points of PG$(p,2^s)$, and the columns are the lines in PG$(p,2^s)$, and it has the following structure:

1. Each row has weight $\rho_r=(2^{ps}-1)/(2^s-1)$.

2. Each column has weight $\rho_c=2^s+1$.

3. Any two columns have at most one 1-component in common.

4. Any two rows have at most one 1-component in common.

The LDPC code with parity-check matrix $\bm{H}_{PG2}(p,s)$ is called a type-2 projective geometry LDPC code, and we denote it by $PG2(p,s)$.

\subsection{Extension of finite geometry LDPC codes by column and row splitting}
\label{IIIC}
A finite geometry LDPC code with $n$ columns and $J$ rows can be extended by splitting each column of its parity-check matrix $H$ into multiple columns. If the splitting is done properly, very good extended finite geometry LDPC codes can be obtained.

Let $\underline{g}_1,\underline{g}_2,...,\underline{g}_n$ be the columns of $\bm{H}$. Let $c_{sp}$ be the column splitting factor, $c_{sp}\in\{1,2,...,\rho_c\}$. Then the column splitting can be done by splitting each $\underline{g}_i$ into $c_{sp}$ columns $\underline{g}_{i,1},\underline{g}_{i,2},...,\underline{g}_{i,c_{sp}}$, and distribute the 1's of the original column among the new columns accordingly, so that the columns $\underline{g}_{i,1},\underline{g}_{i,2},...,\underline{g}_{i,\rho_c-c_{sp}\lfloor\frac{\rho_c}{c_{sp}}\rfloor}$ have weights $\frac{\rho_c}{c_{sp}}+1$, and the other columns have weights $\frac{\rho_c}{c_{sp}}$.

After column splitting, we can proceed with row splitting, that is, determine a row splitting factor $r_{sp}\in\{1,2,...,\rho_r\}$ and follow similarly the process of column splitting.

We denote by $EG1(p,s,c_{sp},r_{sp})$ the LDPC code constructed by an $EG1(p,s)$ LDPC code with column and row splitting factors $c_{sp}$ and $r_{sp}$. The codes $EG2(p,s,c_{sp},r_{sp})$, $PG1(p,s,c_{sp},r_{sp})$, $PG2(p,s,c_{sp},r_{sp})$ are defined similarly.

\section{Simulation Results}
\label{IV}
In this section, we provide simulation results of our QKE protocol with FG codes. We use the same LDPC code for both $C_1$ and $C_2$ in constructing the entanglement-assisted CSS code for our QKE protocol. The channel for quantum communication is assumed to be a depolarizing channel, and the channel error probability $P_e$ in the simulation corresponds to that of the equivalent classical binary-symmetric channel (BSC). We use Monte Carlo simulation with a sample size of $200~000$ for each $P_e$. We allow the SPA decoder to iterate a maximum of $100$ times.

The first group of FG codes we demonstrate is the family of two-dimensional type-1 PG LDPC codes without splitting. These $PG1(2,s,1,1)$ codes require only 1 bit of entanglement per codeword as was proven by Hsieh $et al.$ \cite{Hsieh11} Therefore, it is possible to implement QKE with only 1 bit of preshared secret key. The other families of FG codes we consider are the $EG1(2,5,c_{sp},r_{sp})$ and $PG1(2,5,c_{sp},r_{sp})$ codes.

\subsection{$PG1(2,s,1,1)$ codes}
\label{IVA}
For the $PG1(2,s,1,1)$ codes that require only 1 bit of preshared key, we consider the equivalent BSC bit error probabilities ranging from $0\%$ to $8\%$ in steps of $0.5\%$. Let $[[n,m;c]]$ be the parameters of the entanglement-assisted code, and $R_{net}$ be the original net key rate of QKE using that code; that is, $R_{net}=\frac{m-c}{n}$. This means that the QKE protocol expands a key of length $c$ to a key of length $m$. Table~\ref{tab:tab1} demonstrates all possible $PG1(2,s,1,1)$ codes that have block length $n\leq 10$ $000$.

\begin{table}[h]
\caption{\label{tab:tab1}List of $PG1(2,s,1,1)$ codes that have block length $n\leq 10$ $000$.}
\begin{ruledtabular}
\begin{tabular}{ccc}
$s$ & $[[n,m;c]]$ & $R_{net}$ \\ \hline
2 & $[[21,2;1]]$ & $0.0476$ \\ \hline
3 & $[[73,18;1]]$ & $0.2329$ \\ \hline
4 & $[[273,110;1]]$ & $0.3993$ \\ \hline
5 & $[[1057,570;1]]$ & $0.5383$ \\ \hline
6 & $[[4161,2702;1]]$ & $0.6491$
\end{tabular}
\end{ruledtabular}
\end{table}

\begin{figure}
\includegraphics[width= 8.6cm]{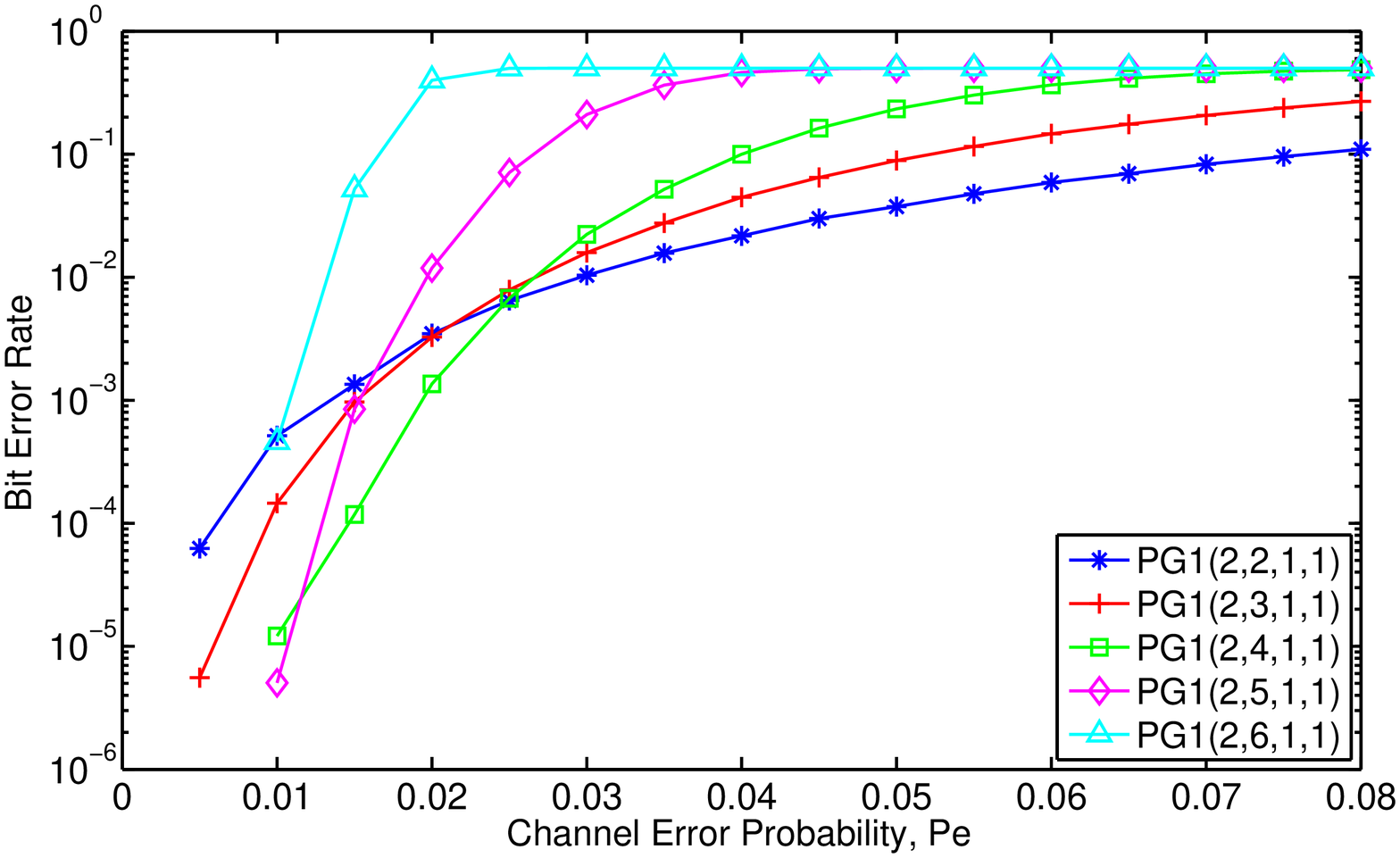}
\caption{(Color online) \label{fig:fig1}Bit error rate of the keys generated by the original QKE protocol with the $PG1(2,s,1,1)$ codes from Table~\ref{tab:tab1}.}
\end{figure}

\begin{figure}
\includegraphics[width= 8.6cm]{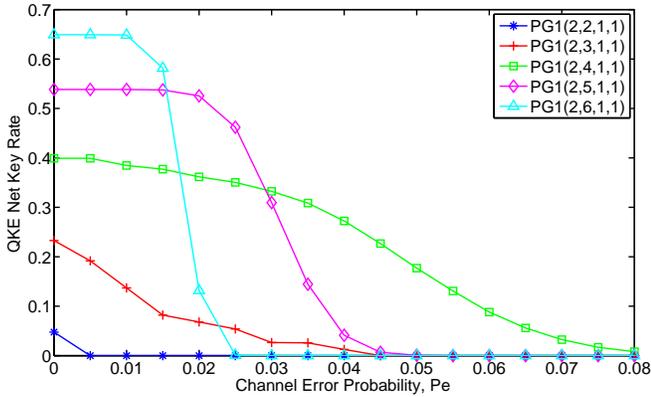}
\caption{(Color online) \label{fig:fig2}Net key rate of the improved QKE protocol with the $PG1(2,5,c_{sp},r_{sp})$ codes from Table~\ref{tab:tab1}. The error threshold is set to $\epsilon=10^{-6}$.}
\end{figure}

In Fig.~\ref{fig:fig1}, we show the QKE performance of the original protocol, in terms of bit error rate, of the codes from Table~\ref{tab:tab1}. In Fig.~\ref{fig:fig2}, we set the generated keys' bit error threshold to $\epsilon=10^{-6}$, and simulate QKE with the improved QKE protocol from Sec.~\ref{II}. We present the performance, in terms of net key rate, using the codes from Table~\ref{tab:tab1}.

The results from Fig.~\ref{fig:fig2} show that it is possible to use just 1 bit of preshared key for QKE even when the channel is moderately noisy up to a bit error probability of $8\%$. In addition, the codes that are considered have reasonable block sizes, and therefore the QKE protocol can be efficiently implemented.

\subsection{$EG1(2,5,c_{sp},r_{sp})$ and $PG1(2,5,c_{sp},r_{sp})$ codes}
\label{IVB}
\begin{figure}
\includegraphics[width= 8.6cm]{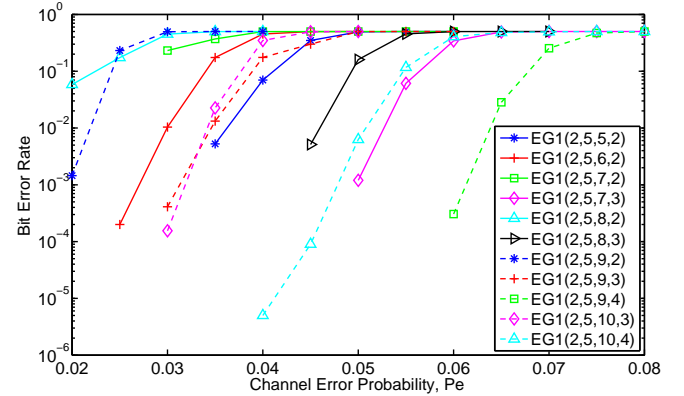}
\caption{(Color online) \label{fig:fig3}Bit error rate of the keys generated by the original QKE protocol with selected codes from $EG1(2,5,c_{sp},r_{sp})$.}
\end{figure}

For the simulation of the $EG1(2,5,c_{sp},r_{sp})$ and $PG1(2,5,c_{sp},r_{sp})$ codes, we consider the equivalent BSC bit error probabilities ranging from $2\%$ to $8\%$ in steps of $0.5\%$. Since many codes perform well when $P_e$ is small, we are mostly interested in codes that have good performance for higher $P_e$, such as might occur in realistic experiments. For a code to serve the purpose of performing key ``expansion," one requires $R_{net}$ to be positive. Table~\ref{tab:tab2} lists all possible $EG1(2,5,c_{sp},r_{sp})$ codes with positive $R_{net}$ that have block length $n\leq 11$ $000$. In Fig.~\ref{fig:fig3}, we show the QKE performance of the original protocol, in terms of bit error rate, of some codes from Table~\ref{tab:tab2}.

\begin{table}[h]
\caption{\label{tab:tab2}List of $EG1(2,5,c_{sp},r_{sp})$ codes with positive net key rates that have block length $n\leq 11$ $000$.}
\begin{ruledtabular}
\begin{tabular}{cccc}
$[[n,m;c]]$ & $c_{sp}$ & $r_{sp}$ & $R_{net}$ \\ \hline
$[[1023,571;32]]$ & $1$ & $1$ & $0.5269$ \\ \hline
$[[2046,452;450]]$ & $2$ & $1$ & $0.0010$ \\ \hline
$[[3069,2045;1022]]$ & $3$ & $1$ & $0.3333$ \\ \hline
$[[4092,3068;1020]]$ & $4$ & $1$ & $0.5005$ \\ \hline
$[[4092,2038;2034]]$ & $4$ & $2$ & $0.0010$ \\ \hline
$[[5115,4091;1022]]$ & $5$ & $1$ & $0.6000$ \\ \hline
$[[5115,3067;2044]]$ & $5$ & $2$ & $0.2000$ \\ \hline
$[[6138,5114;1022]]$ & $6$ & $1$ & $0.6667$ \\ \hline
$[[6138,4090;2044]]$ & $6$ & $2$ & $0.3333$ \\ \hline
$[[7161,6137;1022]]$ & $7$ & $1$ & $0.7143$ \\ \hline
$[[7161,5115;2046]]$ & $7$ & $2$ & $0.4286$ \\ \hline
$[[7161,4092;3069]]$ & $7$ & $3$ & $0.1429$ \\ \hline
$[[8184,7152;1012]]$ & $8$ & $1$ & $0.7502$ \\ \hline
$[[8184,6138;2042]]$ & $8$ & $2$ & $0.5005$ \\ \hline
$[[8184,5115;3067]]$ & $8$ & $3$ & $0.2502$ \\ \hline
$[[8184,4094;4082]]$ & $8$ & $4$ & $0.0015$ \\ \hline
$[[9207,8181;1020]]$ & $9$ & $1$ & $0.7778$ \\ \hline
$[[9207,7161;2046]]$ & $9$ & $2$ & $0.5556$ \\ \hline
$[[9207,6134;3065]]$ & $9$ & $3$ & $0.3333$ \\ \hline
$[[9207,5115;4092]]$ & $9$ & $4$ & $0.1111$ \\ \hline
$[[10230,9202;1018]]$ & $10$ & $1$ & $0.8000$ \\ \hline
$[[10230,8182;2044]]$ & $10$ & $2$ & $0.6000$ \\ \hline
$[[10230,7160;3068]]$ & $10$ & $3$ & $0.4000$ \\ \hline
$[[10230,6132;4086]]$ & $10$ & $4$ & $0.2000$
\end{tabular}
\end{ruledtabular}
\end{table}

\begin{figure}
\includegraphics[width= 8.6cm]{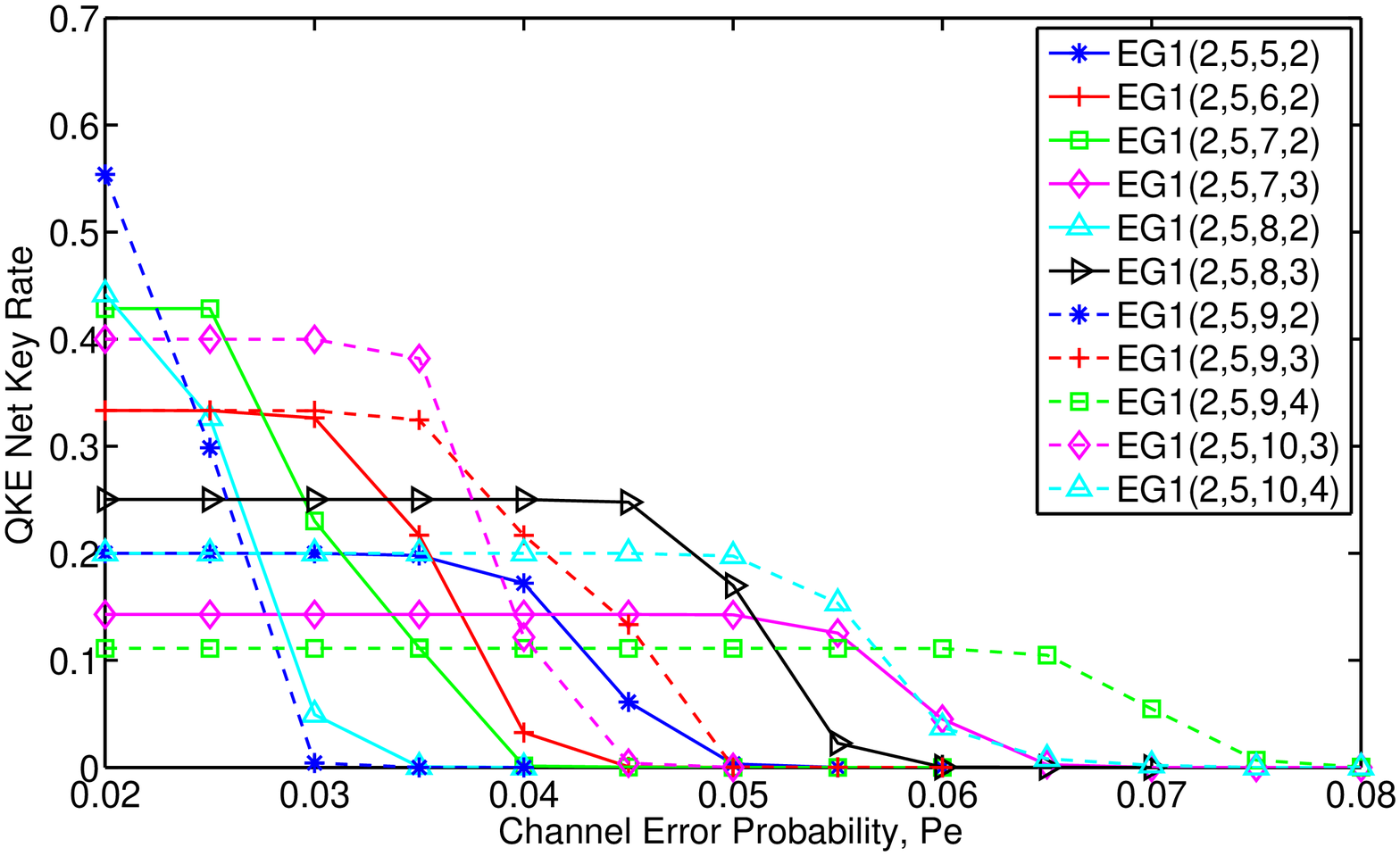}
\caption{(Color online) \label{fig:fig4}Net key rate of the improved QKE protocol with selected codes from $EG1(2,5,c_{sp},r_{sp})$ and error threshold $\epsilon=10^{-6}$.}
\end{figure}

In Fig.~\ref{fig:fig4}, we set the generated keys' bit error threshold to $\epsilon=10^{-6}$, and simulate QKE with the improved QKE protocol from Sec.~\ref{II}. We present the performance, in terms of net key rate, using some codes from Table~\ref{tab:tab2}.

\begin{figure}
\includegraphics[width= 8.6cm]{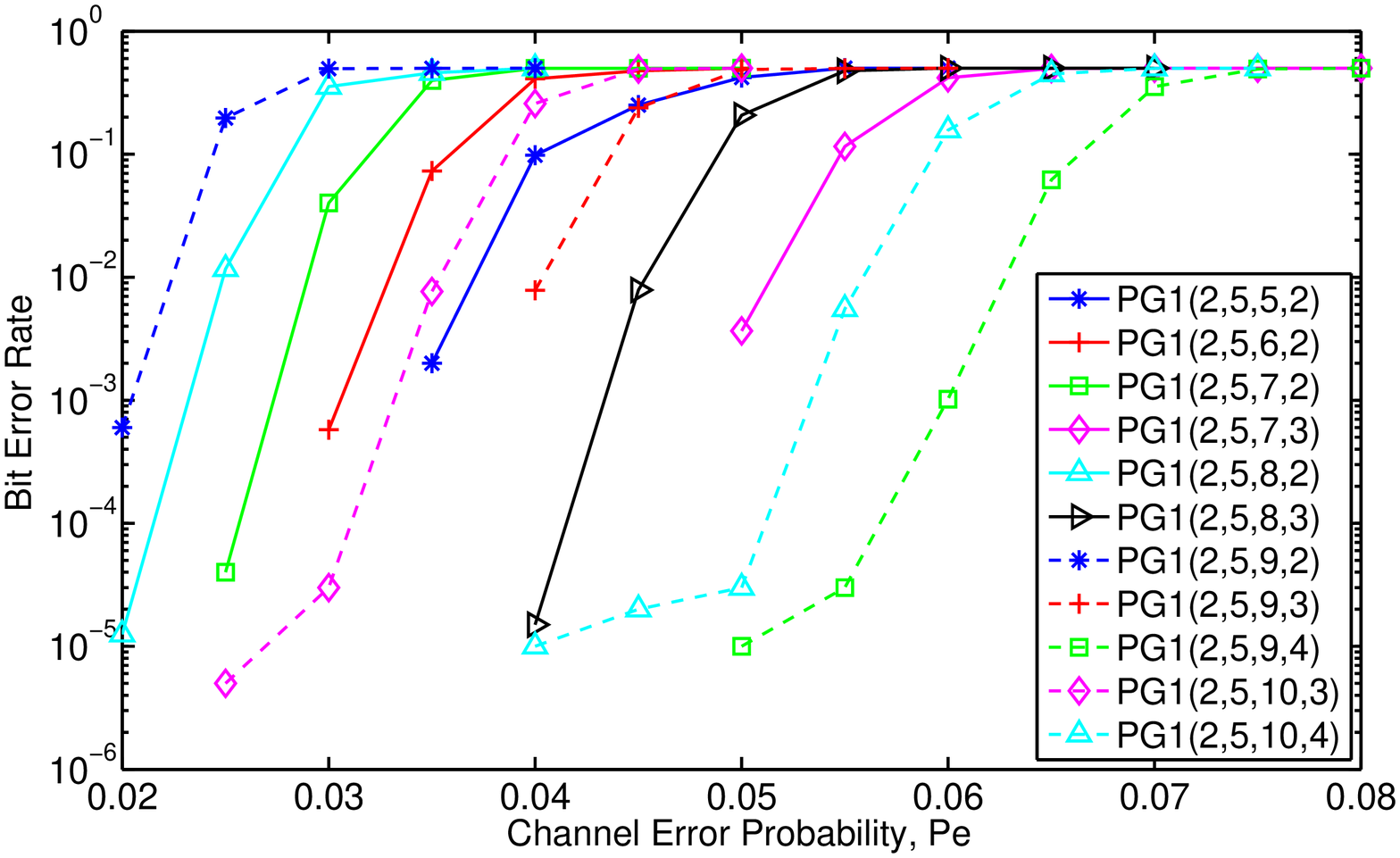}
\caption{(Color online) \label{fig:fig5}Bit error rate of the keys generated by the original QKE protocol with selected codes from $PG1(2,5,c_{sp},r_{sp})$.}
\end{figure}

Table~\ref{tab:tab3} lists all possible $PG1(2,5,c_{sp},r_{sp})$ codes with positive $R_{net}$ that have block length $n\leq 11$ $000$. In Fig.~\ref{fig:fig5}, we present the QKE performance of the original protocol, in terms of bit error rate, of some codes from Table~\ref{tab:tab3}.

\begin{figure}
\includegraphics[width= 8.6cm]{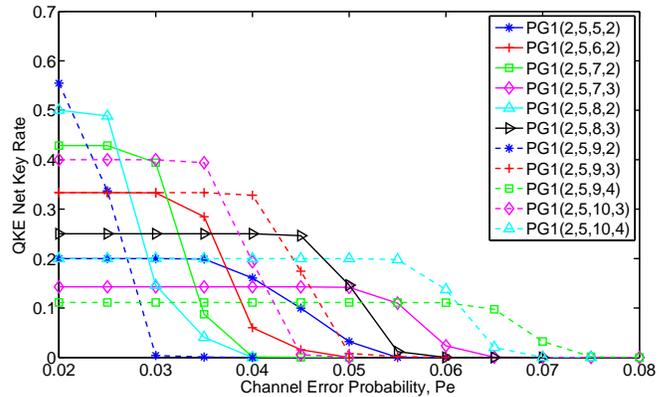}
\caption{(Color online) \label{fig:fig6}Net key rate of the improved QKE protocol with selected codes from $PG1(2,5,c_{sp},r_{sp})$ and error threshold $\epsilon=10^{-6}$.}
\end{figure}

In Fig.~\ref{fig:fig6}, we set the generated keys' bit error threshold to $\epsilon=10^{-6}$ and simulate QKE with the improved QKE protocol proposed in Sec.~\ref{II}. We present the performance, in terms of net key rate, using some codes from Table~\ref{tab:tab3}.

\begin{table}[h]
\caption{\label{tab:tab3}List of $PG1(2,5,c_{sp},r_{sp})$ codes with positive net key rates that have block length $n\leq 11$ $000$.}
\begin{ruledtabular}
\begin{tabular}{cccc}
$[[n,m;c]]$ & $c_{sp}$ & $r_{sp}$ & $R_{net}$ \\ \hline
$[[1057,570;1]]$ & $1$ & $1$ & $0.5383$ \\ \hline
$[[2114,490;488]]$ & $2$ & $1$ & $0.0009$ \\ \hline
$[[3171,2112;1055]]$ & $3$ & $1$ & $0.3333$ \\ \hline
$[[4228,3172;1056]]$ & $4$ & $1$ & $0.5005$ \\ \hline
$[[4228,2114;2112]]$ & $4$ & $2$ & $0.0005$ \\ \hline
$[[5285,4227;1056]]$ & $5$ & $1$ & $0.6000$ \\ \hline
$[[5285,3171;2114]]$ & $5$ & $2$ & $0.2000$ \\ \hline
$[[6342,5284;1056]]$ & $6$ & $1$ & $0.6667$ \\ \hline
$[[6342,4228;2114]]$ & $6$ & $2$ & $0.3333$ \\ \hline
$[[7399,6341;1056]]$ & $7$ & $1$ & $0.7143$ \\ \hline
$[[7399,5285;2114]]$ & $7$ & $2$ & $0.4286$ \\ \hline
$[[7399,4227;3170]]$ & $7$ & $3$ & $0.1429$ \\ \hline
$[[8456,7399;1055]]$ & $8$ & $1$ & $0.7502$ \\ \hline
$[[8456,6342;2112]]$ & $8$ & $2$ & $0.5002$ \\ \hline
$[[8456,5286;3170]]$ & $8$ & $3$ & $0.2502$ \\ \hline
$[[8456,4229;4227]]$ & $8$ & $4$ & $0.0002$ \\ \hline
$[[9513,8455;1056]]$ & $9$ & $1$ & $0.7778$ \\ \hline
$[[9513,7399;2114]]$ & $9$ & $2$ & $0.5556$ \\ \hline
$[[9513,6342;3171]]$ & $9$ & $3$ & $0.3333$ \\ \hline
$[[9513,5284;4227]]$ & $9$ & $4$ & $0.1111$ \\ \hline
$[[10570,9511;1055]]$ & $10$ & $1$ & $0.8000$ \\ \hline
$[[10570,8456;2114]]$ & $10$ & $2$ & $0.6000$ \\ \hline
$[[10570,7399;3171]]$ & $10$ & $3$ & $0.4000$ \\ \hline
$[[10570,6342;4228]]$ & $10$ & $4$ & $0.2000$
\end{tabular}
\end{ruledtabular}
\end{table}

Note that for channel error rates less than $2\%$, we may consider the code $PG1(2,5,9,2)$, which has a net key rate of about $0.5556$. Considering channel error rates much lower than $2\%$, we can use other codes in the family which have even larger net key rates.

\begin{figure}
\includegraphics[width= 8.6cm]{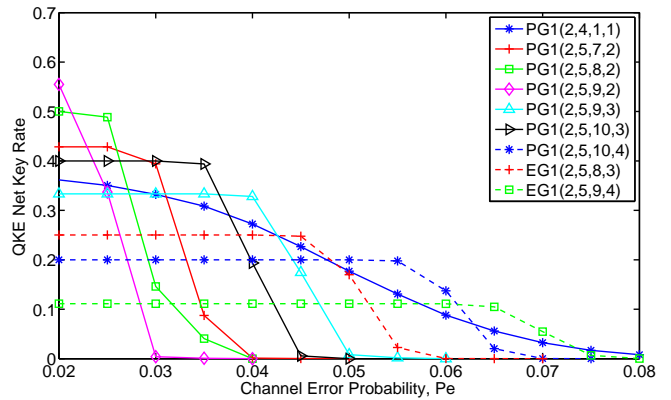}
\caption{(Color online) \label{fig:fig7}Net key rate of the improved QKE protocol with selected codes from both $EG1(2,5,c_{sp},r_{sp})$ and $PG1(2,5,c_{sp},r_{sp})$ that perform well in the various channel error regions.}
\end{figure}

Finally, in Fig.~\ref{fig:fig7}, we set the generated keys' bit error threshold to $\epsilon=10^{-6}$, and we present the QKE net rate using the codes from Tables~\ref{tab:tab1}, ~\ref{tab:tab2}, and \ref{tab:tab3} that perform the best in each channel error region within $2\%$ to $8\%$. As can be seen, quite reasonable key rates can be achieved even for error probabilities above $7\%$.

It is worthwhile comparing our results to the recent work by Elkouss, Leverrier, All\'eaume and Boutros \cite{Elkouss09}. In their work, a set of nine irregular LDPC codes were found for QKD based on the BB84 protocol. With a bit error rate threshold of the generated keys on the same order as ours ($1.5\times 10^{-6}$ in their case), their net key rate performance exceeds ours by roughly $15\%-20\%$ over the same channel error regions. However, this is not too surprising, since they consider LDPC codes with very large block sizes (on the order of $10^6$ bits), while ours have much more modest block sizes (on the order of $10^3$). We believe the sizes of our codes are reasonable for practical use.  Given much greater computing resources for postprocessing, it should be easy to construct very large codes in our family of LDPC codes that would have better net key rates.

\section{Conclusion}
\label{V}
In this paper, we have proposed a protocol for QKE that is an improved version of the protocol proposed by Luo and Devetak. The modifications are done to filter out block errors, which allows us to greatly reduce the bit error rate of QKE with only a small reduction in the net key rate. In addition, we have studied a family of LDPC codes based on finite geometry that are capable of protecting the QKE protocol from errors even when the channel is moderately noisy. The figures in the previous section show clearly which codes one should choose to efficiently expand the keys.

In the near future we will investigate other families of codes for this QKE protocol. The LDPC codes generated by finite geometry are a rich family. Besides the family of FG codes constructed by the method of column and row splitting, we have also examined several codes in a family of quasi-cyclic FG LDPC codes \cite{Chen04,Hsieh09} that perform well for our QKE protocol. Another possible task is to further enhance the QKE protocol. For example, the matrix $\bm{E_1}$ is not unique. If we have a way to search for an $\bm{E_1}$ having density as low as possible, then the block error rate of the code may not affect the bit error rate of the key by as much.

\begin{acknowledgments}
T.A.B. and K.-C.H. would like to acknowledge the Center for High Performance Computing and Communications at the University of Southern California, which provided computing resources. T.A.B. and K.-C.H. also thank Min-Hsiu Hsieh for helpful information and advice, and an anonymous referee for a useful suggestion. This work was supported by NSF Grant No. CCF-0830801.
\end{acknowledgments}

\bibliography{myrefs}

\end{document}